\documentclass[10pt,twocolumn,showpacs,longbibliography,amsmath,amssymb,prd,floatfix,superscriptaddress]{revtex4-2}

\usepackage{graphicx}
\usepackage{dcolumn}
\usepackage{bm}
\usepackage{color}
\usepackage{xcolor} 
\usepackage{txfonts}
\usepackage{microtype}
\usepackage{epstopdf}
\usepackage{pict2e}     
\newcommand\numberthis{\addtocounter{equation}{1}\tag{\theequation}}
\usepackage{tikz-feynman} 
\usepackage{subfigure} 
\definecolor{plotblue}{RGB}{31,119,180}  
\definecolor{plotred}{RGB}{204,85,34}

\begin{document}


\title{Nonlinear Breit–Wheeler Process Driven by Intense Squeezed Light}

\author{Xin Ge}\thanks{These authors contributed equally to this work.}
\affiliation{Department of Physics, Shanghai Normal University, Shanghai 200234, China}
\author{Kai-Hong Zhuang}\thanks{These authors contributed equally to this work.}
\affiliation{Department of Physics, Shanghai Normal University, Shanghai 200234, China}
\author{Pei-Lun He}
\email{peilunhe@sjtu.edu.cn}
\affiliation{State Key Laboratory of Dark Matter Physics, Key Laboratory for Laser Plasmas (Ministry of Education) and School of Physics and Astronomy, Collaborative Innovation Center for IFSA (CICIFSA), Shanghai Jiao Tong University, Shanghai 200240, China}
\author{Yue-Yue Chen}
\email{yueyuechen@shnu.edu.cn}
\affiliation{Department of Physics, Shanghai Normal University, Shanghai 200234, China}
\date {\today}

\begin{abstract}
The nonlinear Breit--Wheeler process is a fundamental phenomenon of strong-field quantum electrodynamics and is usually studied for classically prescribed laser backgrounds.
Here we examine how the statistical properties of a squeezed coherent driving field modify nonlinear Breit--Wheeler pair production.
Using a polarization-resolved Monte Carlo framework with stochastic averaging over the field-amplitude distribution derived from the Husimi $Q$-function, we simulate collisions of $\gamma$ photons with squeezed light and identify clear source-state-dependent modifications of the pair production signal.
{These effects include the smoothing of harmonic structure, the enhancement of higher-order multiphoton channels, and the suppression of the single-laser-photon absorption channel} when stronger-field realizations raise the dressed-mass threshold.
Within the selected spectral window, {the degree of positron polarization increases monotonically with the squeezing parameter}, while the angular distributions broaden as the statistical weight of larger field amplitudes increases.
Our results show that, even at fixed mean electric-field amplitude, the statistical fluctuations inherent to the squeezed coherent state can substantially reshape spectral, angular, and spin-resolved observables in strong-field pair production.
These findings illustrate a direct link between source-state-dependent field statistics and strong-field pair production observables, and provide a theoretical framework for studying how squeezed-state preparation of the driving field can influence high-energy QED processes.
\end{abstract}

\maketitle
\section{Introduction}

The nonlinear Breit--Wheeler (NBW) process converts a high-energy $\gamma$ photon into an electron--positron pair through the absorption of multiple photons from an intense laser field~\cite{titov2018nonlinear,di2016nonlinear}, providing a direct route from light to matter. It is relevant to high-energy astrophysics, laser--plasma interactions, and strong-field physics.
Experimentally, multiphoton pair creation associated with NBW was first observed in the SLAC E-144 experiment in the 1990s~\cite{burke1997positron,bamber1999studies}. In E-144, a 46.6-GeV electron beam collided with a terawatt-class laser pulse (peak intensity $\sim 10^{18}\,\mathrm{W/cm^2}$). The interaction produced high-energy $\gamma$ photons via nonlinear Compton scattering, which then underwent multiphoton pair creation in the same laser field. This provided the first experimental evidence for the NBW mechanism, linked to nonlinear Compton scattering by crossing symmetry, and motivated extensive subsequent studies largely within the classical-background-field paradigm.

The NBW process has since been explored in a wide range of classical and structured backgrounds, including pulsed fields~\cite{tang2021pulse}, dual-pulse configurations with tunable carrier--envelope phase (CEP)~\cite{di2016nonlinear}, and bichromatic drivers~\cite{bulanov2013electromagnetic,meuren2016semiclassical,barbosa2024phase,jiang2024interferences,tang2022fully,blackburn2022higher,blackburn2018nonlinear}. These works highlight the strong sensitivity of NBW to the driving field: by tailoring the laser waveform and parameters, one can control not only the pair yield but also the energy spectra, angular distributions, and spin polarization of the produced leptons.
The polarization of the incident $\gamma$ photon provides an additional control knob. It can strongly modulate both the production rate and the polarization of the outgoing leptons~\cite{wan2020high}, and has been proposed as an indirect probe of vacuum birefringence through NBW-based observables~\cite{borysov2022using}. Furthermore, at moderate field strengths ($a_0 \sim 1$), interference among multiphoton pathways can become pronounced, generating oscillatory structures in angular distributions~\cite{jiang2024interferences}.

With rapid progress in generating intense quantum light~\cite{wang2021generation,slusher1985observation,
wu1986generation,schneider1998generation,furusawa1998unconditional,lam1999optimization,
aoki2006squeezing,mehmet2011squeezed,kumar1984squeezed,yurke1985squeezed},
squeezed states with peak intensities approaching $10^{12}$--$10^{13}\,\mathrm{W/cm^2}$ are now experimentally accessible. Complementary advances have also emerged in the terahertz domain, where ultrafast electro-optic sampling enables access to THz-field vacuum and squeezing fluctuations and supports quantum-enhanced THz detection~\cite{moskalenko2015eos,benea2019vacuum,shields2022eos}.
At these peak intensities, intense squeezed light has been used as a strong-field driver for above-threshold ionization~\cite{fang2023strong,even2024motion,heimerl2024multiphoton,qati1,qati3,mao2025benchmarking},
where photon-number fluctuations can strongly reweight multiphoton channels and imprint heavy-tailed, shot-to-shot yield statistics on the emitted electrons.
Related quantum-statistical effects have also been demonstrated in correlated processes such as nonsequential double ionization~\cite{liu2025atomic} and molecular dissociation~\cite{long2025hydrogen}.
On the radiation side, squeezed-light-driven high-harmonic and attosecond generation~\cite{baerentsen2024squeezed,lemieux2404photon,wang2025attosecond,de2024quantum,rivera2024squeezed,rasputnyi2024high,rivera2025structured}
shows that nonclassical fluctuations can be transferred to the emitted field, modifying harmonic photon statistics and enabling control beyond classical driving.

Although the peak intensities of currently available squeezed states remain below those required for strong-field QED pair creation, it is nevertheless of fundamental interest to understand how nonclassical radiation fields would modify nonlinear QED processes in principle. Because strong-field QED probabilities depend nonlinearly on the background amplitude and are sensitive to higher-order intensity moments, quantum-optical fluctuations can qualitatively influence multiphoton dynamics even at fixed mean intensity.
Initial steps in this direction have recently been taken for nonlinear Compton scattering: squeezed driving fields were shown to reshape the emitted radiation, leading to spectral broadening, enhanced high-frequency emission, and increased directionality relative to coherent states~\cite{khalaf2023compton}, while squeezed vacuum states of the emission modes were shown to significantly enhance or suppress the nonlinear Compton probability even for a classical driving laser~\cite{DiPiazza2026SqueezedVacuum}.  These results demonstrate that extremely nonlinear emission processes can be sensitive to the quantum state of the driving field.
Despite this progress, the impact of quantum-optical states on strong-field pair production remains largely unexplored. In particular, understanding how squeezed light modifies the nonlinear Breit--Wheeler process constitutes a natural extension of these developments.

In this work, we investigate how the quantum-statistical properties of squeezed coherent light modify nonlinear Breit–Wheeler pair production.
Section~II introduces a polarization‑resolved Monte Carlo framework, 
enabling nonperturbative simulations of $\gamma$‑photon interactions with squeezed coherent light. In Sec.~III, we present simulation results for NBW pair production in squeezed light fields, systematically examining the dependence of the energy spectra, spin polarization, and angular distributions of the produced electron--positron pairs on the squeezing parameter. We further analyze the impact of squeezed light on the spectra in a representative low‑$\chi_\gamma$ regime. Finally, we summarize our conclusions in Sec.~IV.



\section{THE THEORETICAL MODEL}

\subsection*{A. Nonlinear Breit--Wheeler probability in a circularly polarized laser field}


The NBW process may be viewed as multiphoton absorption from the laser background:
\begin{equation}
\gamma + n \gamma_L \rightarrow e^+ + e^-,
\end{equation}
where a high-energy probe photon $\gamma$ absorbs $n$ photons $\gamma_L$ from a background laser field to produce a particle–antiparticle pair.
The probability of this process is governed by two Lorentz-invariant parameters:
\begin{equation}
a_0 = \frac{e E_0}{m \omega_L}, \quad 
\eta_\gamma = \frac{k \cdot k_L}{2 m^2} \approx \frac{\omega \omega_L}{m^2},
\end{equation}
where \( E_0 \) and \( \omega_L \) are the peak electric field and frequency of the laser, \( m \) is the electron mass, and \( k \), \( k_L \) denote the four-momenta of the probe and laser photons, respectively. Here, \( a_0 \)  characterizes the laser field strength, while \( \eta_\gamma \) captures the energy of the incoming photon. Natural units with $\hbar=c=1$ are used throughout.

In a circularly polarized monochromatic background, the standard expression for the differential pair production rate per unit time reads \cite{Ivanov:2005polarization}
\begin{equation}\label{LMA}
W = \frac{\alpha m^2 a_0^2}{\omega} 
\sum_{n=n_0}^{\infty} 
\int_{x_n}^{1 - x_n} dx \left( G_{1n} + h_L h_\gamma G_{3n} \right),
\end{equation}
where \( \alpha \) is the fine-structure constant, \( h_\gamma \) and \( h_L \) denote the helicities of the incoming $\gamma$ photon and the laser photons, respectively, \( n_0 = \lceil (1 + a_0^2)/\eta_\gamma \rceil \) represents the minimum number of laser photons required to initiate pair creation, and $x_n=\frac{1}{2}(1-\sqrt{1-\frac{1}{u_n}})$ with \( u_n = \frac{n\eta_\gamma}{1 + a_0^2} \) denotes the maximum value of \( u \) for a given harmonic number \( n \).
If the polarization of the positron is also resolved, the rate generalizes to
\begin{multline}\label{LMA_spin}
W = \frac{\alpha m^2 a_0^2}{2\omega} 
\sum_{n=n_0}^{\infty} 
\int_{x_n}^{1 - x_n} dx \Big[ 
G_{1n} + h_L h_\gamma G_{3n} \\
+ \bar{h}_e \left( h_L G_{2n} + h_\gamma G_{4n} \right) 
\Big],
\end{multline}
where \( \bar{h}_e \) is the helicity of the emitted positron. The coefficient functions are
\begin{equation}
\begin{aligned}
G_{1n} &= \frac{1}{a_0^2} J_n^2 + \frac{1}{2}(2u - 1)\left(J_{n-1}^2 + J_{n+1}^2 - 2 J_n^2 \right), \\
G_{2n} &= (1 - 2x) 2u \left( \frac{1}{2} - \frac{u}{u_n} \right) \left( J_{n-1}^2 - J_{n+1}^2 \right), \\
G_{3n} &= -(2u - 1) \left( \frac{1}{2} - \frac{u}{u_n} \right) \left( J_{n-1}^2 - J_{n+1}^2 \right), \\
G_{4n} &= \frac{1}{x a_0^2} J_n^2 - u (1 - 2x) \left( J_{n-1}^2 + J_{n+1}^2 - 2 J_n^2 \right),
\end{aligned}
\end{equation}
where \( J_n \) is the Bessel function of the first kind. The argument is given by
\begin{equation}
z_n = 2n \frac{a_0}{\sqrt{1 + a_0^2}} \sqrt{ \frac{u}{u_n} \left( 1 - \frac{u}{u_n} \right) },
\end{equation}
where \( u = \frac{1}{4x(1 - x)} \) with \( x = \frac{p\cdot k_L}{k\cdot k_L}\;\approx\;\frac{\varepsilon}{\omega} \).
These Bessel functions capture the multiphoton interference and harmonic structures intrinsic to the nonlinear interaction.

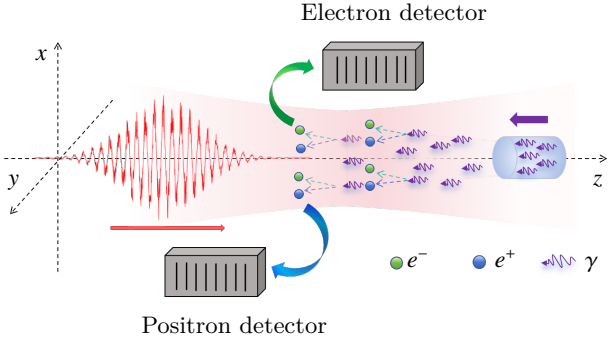
\begin{figure}
    \centering
    \includegraphics[width=0.45\textwidth]{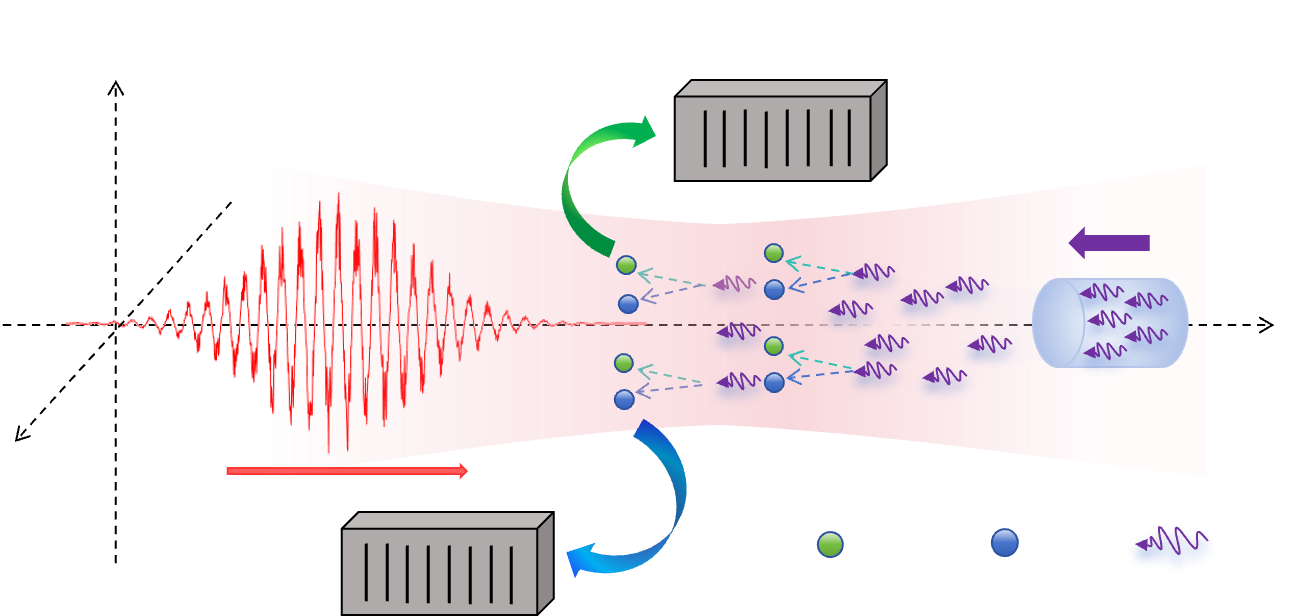}
     \begin{picture}(300,20)
         \put(20,110){$x$}
         \put(10,63){$y$}   
         \put(230,63){$z$}      
        \put(120,125){Electron detector}
         \put(60,7){Positron detector}             
         \put(160,32){$e^-$}
         \put(193,32){$e^+$}
         \put(227,32){$\gamma$}
    \end{picture}
\caption{Schematic of nonlinear Breit--Wheeler pair production driven by a quantum squeezed‑light pulse.
A high‑energy $\gamma$‑ray beam (purple, propagating along $-z$) collides head‑on with an intense, circularly polarized squeezed laser pulse (red, propagating along $+z$).
Within the focal overlap, the simultaneous absorption of multiple laser photons converts the incoming $\gamma$ photon into an $e^{+}e^{-}$ pair.}
    \label{Fig:scheme}
\end{figure}

\subsection*{B. Squeezed coherent light}


In conventional studies of the NBW process, the background laser field is typically treated as a classical coherent state, with quantum fluctuations neglected. A coherent state is defined as the vacuum acted upon by the displacement operator,
\begin{equation}
|\alpha\rangle = D(\alpha)\,|0\rangle, \qquad
D(\alpha)=\exp\!\left(\alpha a^\dagger-\alpha^* a\right),
\end{equation}
where $\alpha\in\mathbb{C}$ characterizes the complex field amplitude, and $a^\dagger$ and $a$ are the photon creation and annihilation operators satisfying the canonical commutation relation $[a,a^\dagger]=1$.
For coherent light polarized along the unit vector $\hat{\mathbf{e}}$, the classical laser field is recovered as the expectation value of the field operator in the coherent state,
\begin{equation}
\label{field1}
\langle \alpha | \hat{\mathbf{E}}(\varphi) | \alpha \rangle
=
2\epsilon_V \bigl[\alpha_1 \cos\varphi + \alpha_2 \sin\varphi \bigr]\hat{\mathbf{e}},
\end{equation}
where $\varphi=\omega t-kz$ denotes the laser phase, and the subscripts $1,2$ label the real and imaginary parts of $\alpha$, respectively. Here
$\epsilon_V=\sqrt{\frac{\hbar\omega}{2\epsilon_0 V}}$ is the single‑photon field amplitude, with $\hbar$ the reduced Planck constant, $\epsilon_0$ the vacuum permittivity, and $V$ the quantization volume of the laser pulse~\cite{qV}.
The two orthogonal quadrature components of the electromagnetic field,
\begin{equation}
X_1=\tfrac{1}{2}(a+a^\dagger), \qquad
X_2=\tfrac{1}{2i}(a-a^\dagger),
\end{equation}
exhibit equal standard deviations in a coherent state,
$\Delta X_1=\Delta X_2=\tfrac{1}{2}$, thereby saturating the minimum uncertainty relation
$\Delta X_1\,\Delta X_2=\tfrac{1}{4}$. The photon‑number distribution is Poissonian, with mean photon number
$\langle n\rangle = |\alpha|^2$ .

In contrast, a nonclassical squeezed state is characterized, at the operator
level, by reduced quantum fluctuations in one field quadrature accompanied by
enhanced fluctuations in the conjugate quadrature. A squeezed vacuum state is
generated from the vacuum by the squeezing operator,
\begin{equation}
|\zeta\rangle=S(\zeta)|0\rangle,\qquad
S(\zeta)=\exp\!\left[\tfrac{1}{2}\!\left(\zeta a^2-\zeta^*(a^\dagger)^2\right)\right],
\end{equation}
where {$\zeta=re^{i\phi}$} is the complex squeezing parameter, with $r$
denoting the squeezing strength and {\(\phi\)} the squeezing angle. The mean
photon number of a squeezed vacuum state is
\begin{equation}
\langle n\rangle=\sinh^2 r.
\end{equation}

To incorporate both classical excitation and quantum squeezing, we consider
the squeezed coherent state, defined as a coherent displacement of a
squeezed vacuum,
\begin{equation}
|\beta,\zeta\rangle = D(\beta)\,S(\zeta)\,|0\rangle .
\end{equation}
For this state, the mean photon number naturally separates into classical and
quantum contributions,
\begin{equation}
\langle n\rangle = |\beta|^2 + \sinh^2 r,
\end{equation}
where $|\beta|^2$ represents the photon number associated with the coherent
excitation, while $\sinh^2 r$ originates from squeezing-induced quantum
fluctuations. In the limit $r\to0$, the squeezed coherent state reduces to a
coherent state, and its statistical properties revert to the Poissonian
statistics of classical light.

The statistical properties of a squeezed coherent state can be conveniently described by its Husimi $Q$‑function,
\begin{equation}
Q_r(\alpha,\phi)
=
\frac{\exp\!\left[-|\alpha-\beta|^2\right]}{\pi\cosh r}
\exp\!\left[
-\frac{1}{2}\tanh r\,(\alpha-\beta)^2 e^{-i\phi}
+ \text{c.c.}
\right],
\end{equation}
where $\beta$ denotes the coherent amplitude of the displaced squeezed state.

For {\(\phi=0\)}, the Husimi $Q$‑function reads
\begin{equation}
Q_{r}(\alpha)=
\frac{1}{\pi \cosh r}
\exp\!\left[
-\frac{2(\alpha_1-\beta_1)^2}{1+e^{-2r}}
-\frac{2(\alpha_2-\beta_2)^2}{1+e^{2r}}
\right].
\label{eq:14}
\end{equation}
Choosing the phase reference of the coherent field such that $E_{\beta 1}=0$ and $E_{\beta 2}=E_\beta$, we focus on phase squeezing and implement it through the macroscopic‑limit electric‑field quasiprobability distribution.  
In the macroscopic limit $V\to\infty$, while keeping the classical field amplitude finite, the corresponding electric‑field quasiprobability distribution for a phase‑squeezed state $|\beta,r\rangle$ can be approximated as
\begin{equation}
\tilde{Q}(E_{\alpha}) \simeq
\frac{1}{\sqrt{2\pi}\,E_{\mathrm{vf}}}
\exp\!\left[
-\frac{(E_{\alpha 2}-E_{\beta 2})^{2}}{2E_{\mathrm{vf}}^{2}}
\right]
\delta(E_{\alpha 1}-E_{\beta 1}),
\label{eq:15}
\end{equation}
where $E_{\mathrm{vf}} \approx \epsilon_V e^{r}$ denotes the characteristic electric‑field fluctuation amplitude entering the statistical ensemble of classical field realizations.  
By absorbing the single‑photon amplitude into the macroscopic normalization and taking the limit in Eq.~(\ref{eq:15}), the expression for the expectation value of the electric field changes from Eq.~(\ref{field1}) to
\begin{equation}
\mathbf{E}_\alpha(\varphi) 
=
\bigl[E_{\alpha1} \cos\varphi + E_{\alpha2} \sin\varphi \bigr]\hat{\mathbf{e}} .
\end{equation}

{We also note that the squeezing angle \(\phi\) determines the orientation of the
noise ellipse in quadrature space and can redistribute fluctuations between
different quadratures. In the present Husimi-\(Q\) sampling framework, however,
the squeezed field enters the NBW probability through the effective field
amplitude. For the plane-wave background considered here, the coarse-grained
observables, such as the positron yield, spectrum, angular distribution, and spin
polarization, are governed mainly by the local photon quantum parameter
\(\chi_\gamma\), rather than by the absolute optical phase. Therefore, varying
\(\phi\) mainly changes the effective fluctuation strength and does not
introduce a qualitatively new phase-coherent contribution within the present
approximation.
}

\begin{figure*}
\centering
\includegraphics[width=0.98\textwidth]{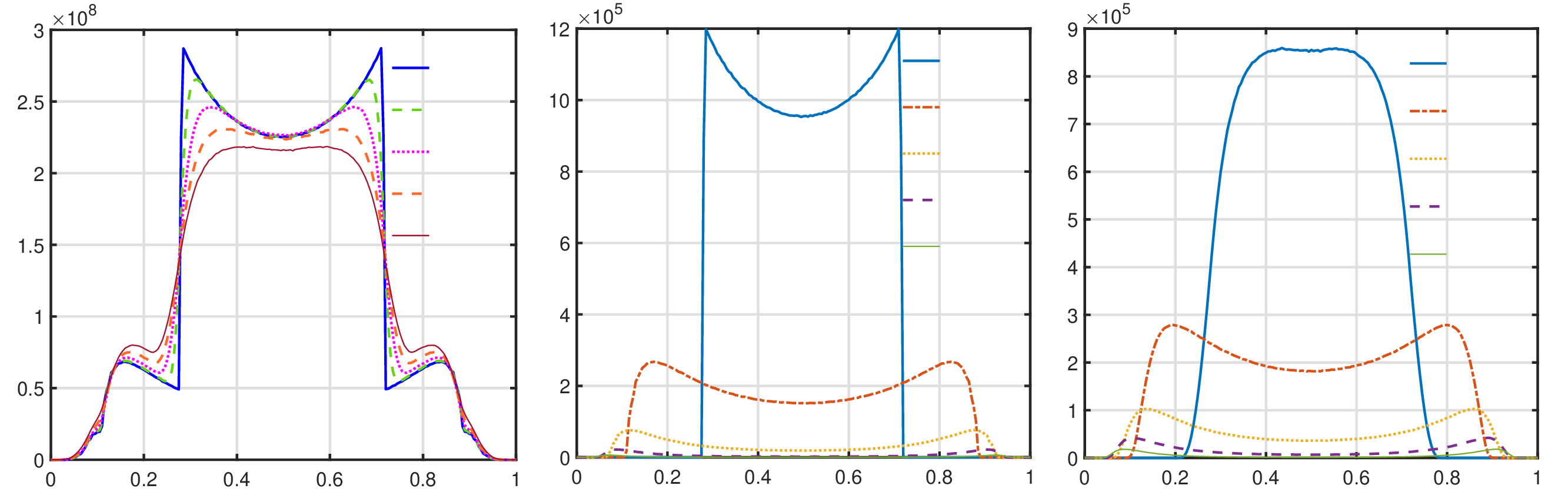}
 \begin{picture}(300,20)
        \put(-77,155){{(a)}}
        \put(91,155){{(b)}}
        \put(255,155){{(c)}}
        
        \put(40,157){\scriptsize ${\rho} = 0.0$}
        \put(40,144){\scriptsize ${\rho} = 0.1$}
        \put(40,132){\scriptsize ${\rho} = 0.2$}
        \put(40,118){\scriptsize ${\rho} = 0.3$}
        \put(40,104) {\scriptsize ${\rho} = 0.4$}
        
     \put(203,160){\scriptsize $n_{\mathrm{pho}} = 1$}
\put(203,145){\scriptsize $n_{\mathrm{pho}} = 2$}
\put(203,130){\scriptsize $n_{\mathrm{pho}} = 3$}
\put(203,116){\scriptsize $n_{\mathrm{pho}} = 4$}
\put(203,101){\scriptsize $n_{\mathrm{pho}} = 5$}

    \put(365,159){\scriptsize $n_{\mathrm{pho}} = 1$}
\put(365,144){\scriptsize $n_{\mathrm{pho}} = 2$}
\put(365,128){\scriptsize $n_{\mathrm{pho}} = 3$}
\put(365,113){\scriptsize $n_{\mathrm{pho}} = 4$}
\put(365,99){\scriptsize $n_{\mathrm{pho}} = 5$}

        \put(-105,90){\rotatebox{90}{\small $dN/d\delta$}}  
        \put(-10,10){\small $\delta$}              
     
        \put(158,10){\small $\delta$}            

        \put(322,10){\small $\delta$}                                  
        \end{picture}
    \caption {%
(a) Positron energy spectra \( dN/d\delta \) as a function of the normalized energy fraction \( \delta \equiv \varepsilon_+ / \omega \), 
where \(\omega = 250\,\mathrm{GeV}\) and \(\varepsilon_+\) denote the incident photon energy and the produced positron energy, respectively.
The different curves correspond to reduced squeezing parameters \( {\rho} = 0.0 \) (blue), \( 0.1 \) (green), \( 0.2 \) (magenta), \( 0.3 \) (orange), and \( 0.4 \) (red).  
Positron spectra decomposed into multiphoton absorption channels in the nonlinear Breit--Wheeler process: (b) coherent light (\( {\rho} = 0.0 \)); (c) squeezed light (\( {\rho} = 0.4 \)). Each curve corresponds to the absorption of \( n_{\mathrm{pho}} = 1 \) (blue), 2 (orange), 3 (yellow), 4 (magenta), and 5 (green) laser photons.  
}

    \label{fig:spectrum_nbw}
\end{figure*}
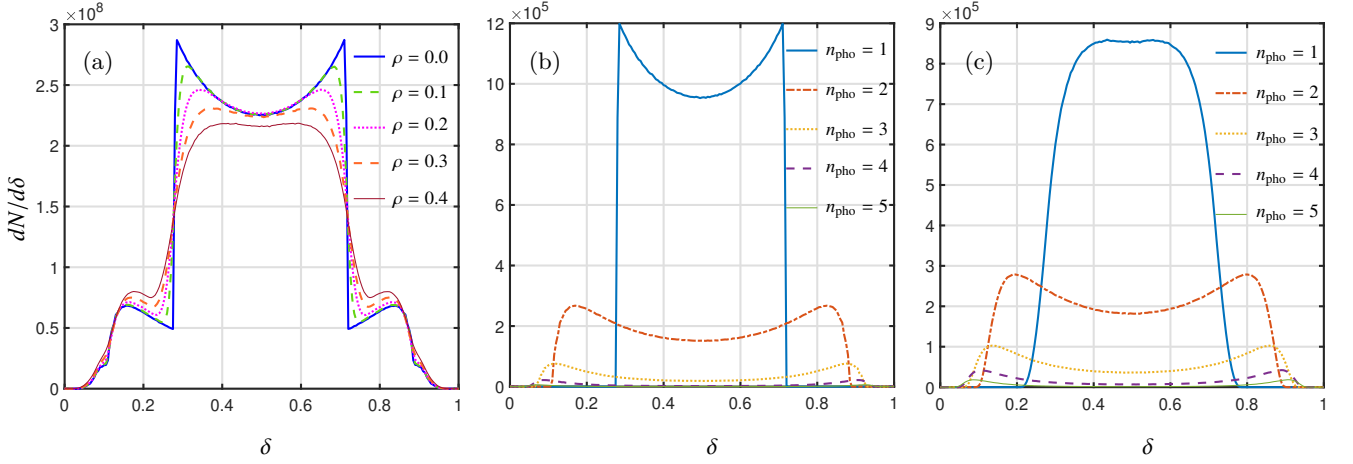

\subsection*{C. Simulation implementation for the nonlinear Breit--Wheeler process driven by quantum light}

For intense quantum light interacting with matter, the evaluation of the nonlinear Breit--Wheeler pair‑production probability reduces to a classical stochastic average over the Husimi $Q$‑function [see Eq.~(\ref{eq:spectrum_conv})].  
This approach has previously been employed for nonlinear Compton scattering in squeezed backgrounds \cite{khalaf2023compton}. Its validity in the macroscopic photon‑number regime has been benchmarked against first‑principles time‑dependent Schr\"odinger  equation calculations with quantized radiation fields, showing quantitative agreement \cite{mao2025benchmarking}.  
Here, we extend the same formalism to the nonlinear Breit--Wheeler process, since the background field considered also contains a macroscopic number of photons.

Accordingly, in the Monte Carlo implementation, the fluctuating quadrature of the background field is modeled as
\begin{equation}
\delta E_{\alpha 2} = E_{\mathrm{vf}}\,\xi,
\qquad
\xi\sim\mathcal{N}(0,1),
\label{eq:13}
\end{equation}
while the orthogonal quadrature is kept fixed. For each $\gamma$ photon, we draw an independent random variable \(\xi\) at the start of the interaction, which determines the peak field amplitude entering the strong-field QED rates:
\begin{equation}
E_\alpha
=
E_\beta\left(1+{\rho}\,\xi\right),
\label{eq:16}
\end{equation}
where the dimensionless reduced squeezing parameter is defined as
\begin{equation}
{\rho} \equiv \frac{E_{\mathrm{vf}}}{E_\beta}.
\end{equation}
In this way, each $\gamma$ photon is propagated in a background field with an independently sampled peak amplitude, determined by its own realization of \(\xi\). 

The occurrence of pair production is likewise determined stochastically using the standard Monte Carlo algorithm. 
At each time step, the pair production probability, \(P=W\,\delta t\), is evaluated using Eq.~(\ref{LMA}), and a Monte Carlo criterion is then applied to determine whether pair production occurs.
A uniformly distributed random number \( R \in [0,1] \) is drawn and compared with the instantaneous interaction probability \( P \). 
If \( R > P \), the photon propagates without interacting; 
otherwise, a pair-creation event is registered.
When an event is accepted, the absorbed-photon number \( n \) and the positron energy fraction \( x \) are sampled from the differential rate in Eq.~(\ref{LMA}). 
The emission angle \( \theta \) of the positron is then fixed by the chosen \( n \) and \( x \) following \cite{Yokoya2017_CAIN}
\begin{equation}\label{theta}
\theta_{e^+} = \frac{m\sqrt{4{\eta_\gamma} nx(1-x)-(1+a_0^2)}}{\omega(1-x)}.
\end{equation}
The positron’s longitudinal spin polarization is obtained from the spin-resolved rate [Eq.~(\ref{LMA_spin})], yielding the average value
\begin{equation}
h_{e} = \frac{h_{L} G_{2n} + h_{\gamma} G_{4n}}
              {G_{1n} + h_{L} h_{\gamma} G_{3n}}.
\end{equation}
For every generated electron–positron pair, the simulation records the energy, momentum, and spin polarization.
This numerical framework integrates quantum fluctuations from squeezed light into the conventional QED simulations, enabling accurate modeling of strong-field QED in nonclassical light backgrounds. 

{We now clarify the validity regime and limitations of the Husimi-\(Q\) averaging
method. This treatment is a semiclassical, coarse-grained description of the
squeezed laser mode, applicable for macroscopic field occupation
\(|\beta|^2\gg1\), moderate squeezing, and observables averaged over many
pair-production events. In this regime, the laser mode is represented by an
ensemble of classical field amplitudes sampled from the positive Husimi
\(Q\)-function, and the nonlinear Breit--Wheeler probability is evaluated for
each sampled realization. The method captures the modified quadrature-fluctuation
statistics of the squeezed state, but does not retain quantum coherence between
different coherent-state components \cite{mao2025benchmarking,zhou2026attosecond}. Thus, the reported effects should be
interpreted as consequences of squeezed-state fluctuation statistics in the
large-photon-number limit, rather than as phase-coherent interference effects of
the laser mode. Corrections associated with operator ordering, sub-vacuum
quadrature fluctuations, or temporal correlations may become relevant for small
photon numbers, very strong squeezing, or phase- and time-resolved observables.}

\section{Simulation Results and Analysis}
In this section, we investigate how the squeezing parameter influences key properties of the produced positrons, including their yield, spin polarization, and angular distribution. The interaction setup consists of a monochromatic, circularly polarized laser pulse with intensity parameter $a_0^2=0.2$, helicity $h_L=-1$, and wavelength $\lambda=800~\mathrm{nm}$, propagating along the $+z$ direction. The laser collides head‑on with a beam of $\gamma$‑ray photons of energy $\omega=250~\mathrm{GeV}$, propagating along the $-z$ direction. The number of incident photons is $N_\gamma=2.5\times10^8$, and the interaction duration is $300T$, where $T$ denotes the laser period.
This configuration corresponds to the perturbative multiphoton regime with {$\chi_\gamma=2a_0\eta_\gamma\sim1$}. 
{Although recent progress in bright squeezed light is encouraging~\cite{Kern2026,rasputnyi2024high}, the parameter regime considered here should be viewed as a long-term goal for future squeezed-light-driven NBW experiments rather than a near-term configuration.}
An illustration of the setup is shown in Fig.~\ref{Fig:scheme}.

\subsection*{A. Positron spectrum}

The resulting positron energy spectra for different squeezing parameters are shown in Fig.~\ref{fig:spectrum_nbw}.
For coherent light ($r=0$), the positron spectrum exhibits pronounced harmonic peaks [Fig.~\ref{fig:spectrum_nbw}(a)], reflecting discrete multiphoton absorption in strong-field QED. As the {reduced squeezing parameter $\rho$} increases, corresponding to enhanced quantum fluctuations, the spectrum becomes progressively smoother and broader. In particular, the yield in the intermediate energy range $\delta\in[0.3,0.7]$ is suppressed, while the spectral weight at both low- and high-energy edges is enhanced, indicating a redistribution of pair production probability induced by the modified field statistics.

To elucidate the origin of these spectral changes, Figs.~\ref{fig:spectrum_nbw}(b,c) decompose the positron spectra into photon-absorption channels ($n_{\text{pho}}=1$--$5$)~\cite{khalaf2023compton,wang2025attosecond},
\begin{equation}
\frac{dN}{d\delta}\bigg|_{r}
=\sum_{n=1}^{5}\!\int\! dE_\alpha\, \tilde{Q}(E_\alpha)\, W_n(\delta; E_\alpha),
\label{eq:spectrum_conv}
\end{equation}
where $\delta=\varepsilon_+/\omega$ denotes the normalized positron
energy fraction.
Here $W_n(\delta;E_\alpha)$ denotes the channel‑resolved nonlinear Breit--Wheeler spectrum corresponding to the absorption of $n$ laser photons. Its explicit expression is given in Eq.~(3) and is evaluated at a fixed field amplitude, with the intensity parameter satisfying $a_0 \propto E_\alpha$.
For coherent light, $\tilde{Q}(E_\alpha)=\delta(E_{\alpha}-E_\beta)$, and Eq.~(\ref{eq:spectrum_conv}) reduces to a discrete sum over channels evaluated at a single intensity parameter.
In the coherent case, the single-photon channel ($n_{\text{pho}}=1$) dominates the spectrum, displaying a broad flat-top structure with side maxima near $\delta\simeq0.2$ and $0.8$ and a shallow central dip. Higher-order channels contribute progressively less and extend toward the low- and high-energy edges, signaling the onset of multiphoton processes.

\begin{figure}
    \centering

    \includegraphics[width=0.45\textwidth]{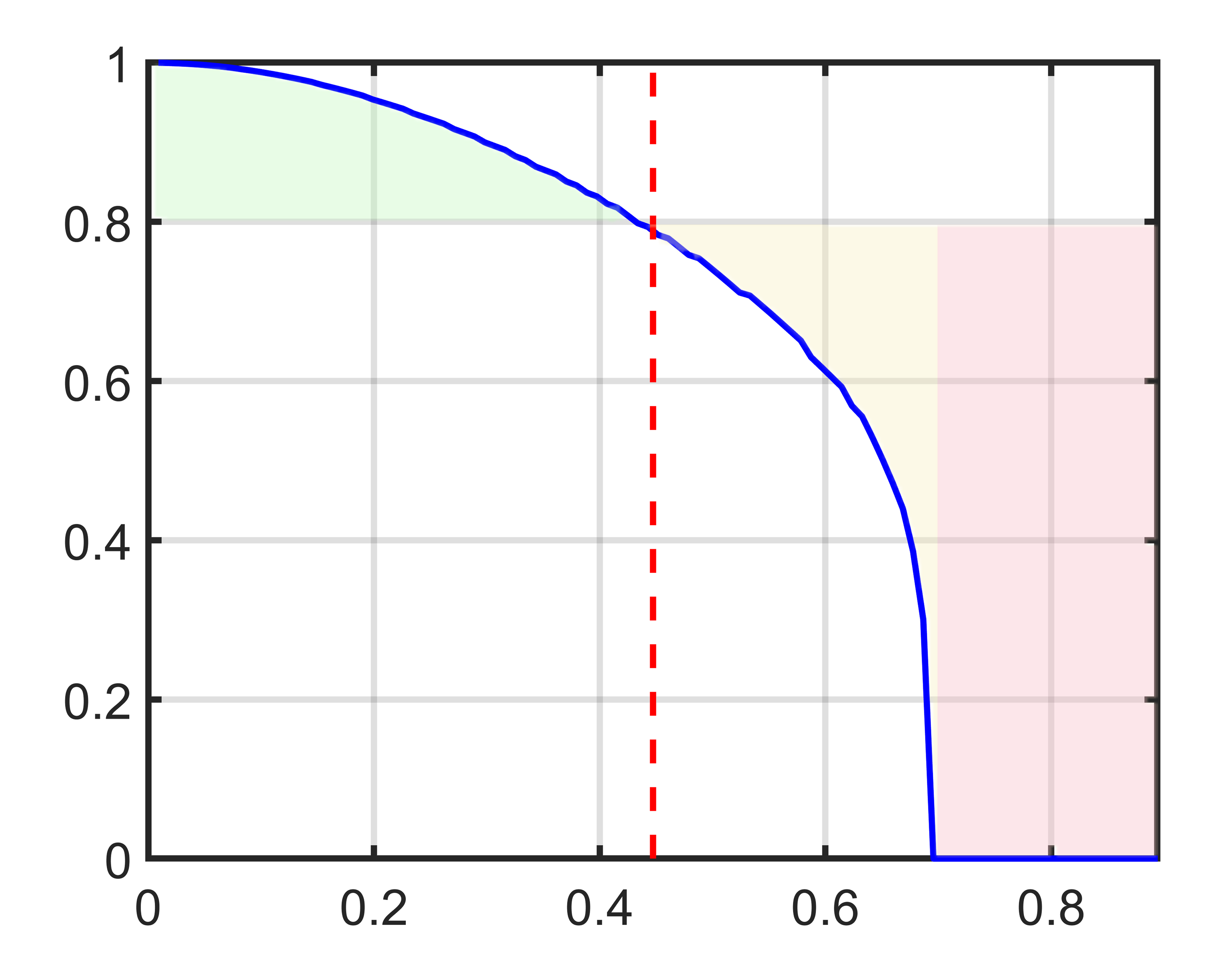}
   \begin{picture}(300,20)
       \put(60,167){$(i)$}
       \put(155,144){$(ii)$}
       \put(195,115){$(iii)$}
        \put(5,115){\rotatebox{90}{\small $R$}}  
        \put(130,12){\small $a_0$}              
         \put(135,45){\small $a_\beta$}  
    \end{picture}
\caption{Relative weight of the single-photon channel ($n_{\rm pho}=1$),
$R \equiv W_{n=1}/\sum_n W_n$, as a function of the field amplitude $a_0$ (solid blue line).
The dashed red line marks the coherent-driving case, $a_0=a_\beta$ [$a^2_\beta=0.2$].
The shaded bands indicate three regimes: (i) an enhancement region with $R>0.8$,
(ii) a reduction region with $R<0.8$, and
(iii) a kinematically forbidden region where $R=0$, i.e., the $n_{\rm pho}=1$ contribution vanishes.}
    \label{fig3}
\end{figure}
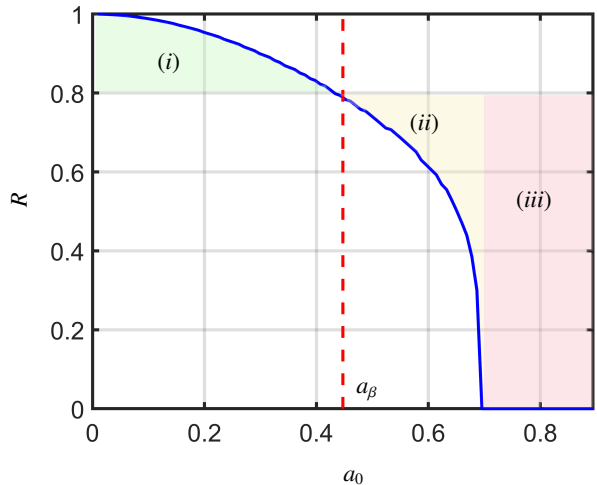

When squeezing is introduced, the single-photon channel ($n_{\rm pho}=1$) remains dominant; however, its sharp spectral edges are noticeably smoothed and its positron yield decreases. This reduction stems from a pronounced drop of the channel weight
\(
R \equiv W_{1}\big/\sum_{n} W_{n}
\)
at larger field amplitudes $a_0$. For coherent driving at $a_0=a_\beta$ [$a^2_\beta=0.2$], the relative contribution of the single-photon channel is $R\simeq 0.8$. In a squeezed state, amplitude fluctuations make $a_0$ a stochastic variable distributed around $a_\beta$. As a result, the $n_{\rm pho}=1$ channel is enhanced for realizations with $a_0<a_\beta$ but suppressed for $a_0>a_\beta$ (see Fig.~\ref{fig3}). Crucially, the suppression at higher $a_0$ is much stronger than the enhancement at lower $a_0$, and the $n_{\rm pho}=1$ channel becomes kinematically forbidden for $a_0\gtrsim 0.7$. This occurs because increasing $a_0$ raises the dressed mass $m_*$ of the produced leptons and thereby elevates the NBW threshold \cite{ritus1985quantum}. Averaging over the squeezed-state amplitude distribution therefore shifts statistical weight into the high-$a_0$ region where $R$ is strongly reduced (or vanishes), yielding a net suppression of the positron yield of $n_{\rm pho}=1$ channel relative to the coherent case. 

By contrast, the higher-order channels ($n_{\text{pho}}=2$--$5$) are systematically enhanced across the spectrum, with particularly strong amplification near the spectral boundaries. This behavior can be traced back to the broadened field-amplitude distribution associated with squeezed-state driving, which generates an ensemble of field strengths rather than a fixed value. In our Monte Carlo implementation, this ensemble is realized by sampling the intensity parameter $a_0$ from the quasiprobability distribution $\tilde Q(a_0)$, whose width scales as $\Delta a_0\propto{\rho}$.
In the parameter regime considered here, the partial NBW rate for absorbing $n$ laser photons approximately follows the scaling $W_n(a_0)\propto a_0^{2n}$, and thus behaves as a monotonically increasing and locally convex function of $a_0$, with a progressively steeper logarithmic slope for larger $n$. Consequently, Jensen’s inequality~\cite{tsang_quantum_optics} implies
\begin{equation}
\langle W_n(a_0)\rangle_{\tilde Q}
>
W_n\!\left(\langle a_0\rangle_{\tilde Q}\right),
\end{equation}
with the enhancement becoming progressively stronger for higher-order channels. 
Physically, this reflects the heightened sensitivity of large-$n$ processes to comparatively rare large-amplitude realizations in the tail of $\tilde Q(a_0)$~\cite{spasibko2017multiphoton}.
Note that the Jensen-type enhancement applies only to channels that remain kinematically allowed over the sampled $a_0$ range. The $n_{\rm pho}=1$ channel is special because it is cut off for $a_0 \gtrsim 0.7$, so the heavy tail of $\tilde{Q}(a_0)$ contributes primarily zero weight to $W_1$ while still boosting $W_{n\ge 2}$.
This theoretical expectation is confirmed by the spectral decomposition shown in Fig.~\ref{fig:spectrum_nbw}(c), where all channels with $n\geq2$ exhibit order-dependent enhancement relative to the coherent case (${\rho}=0$).


 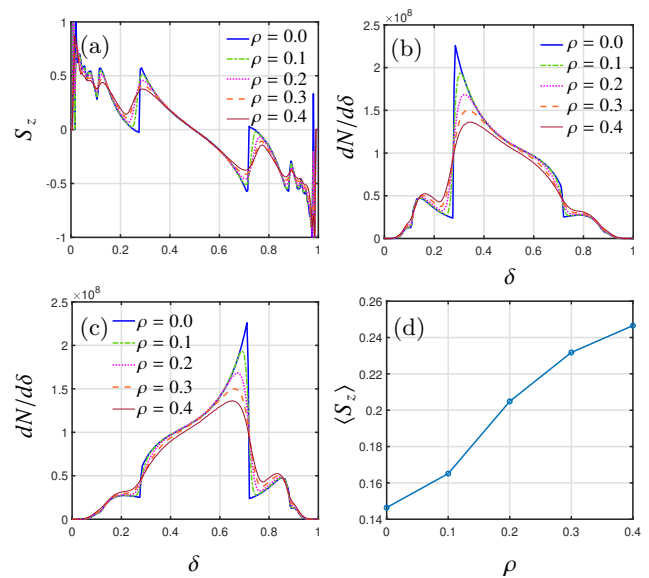
\begin{figure}[b]
    \centering
    \includegraphics[width=0.95\linewidth]{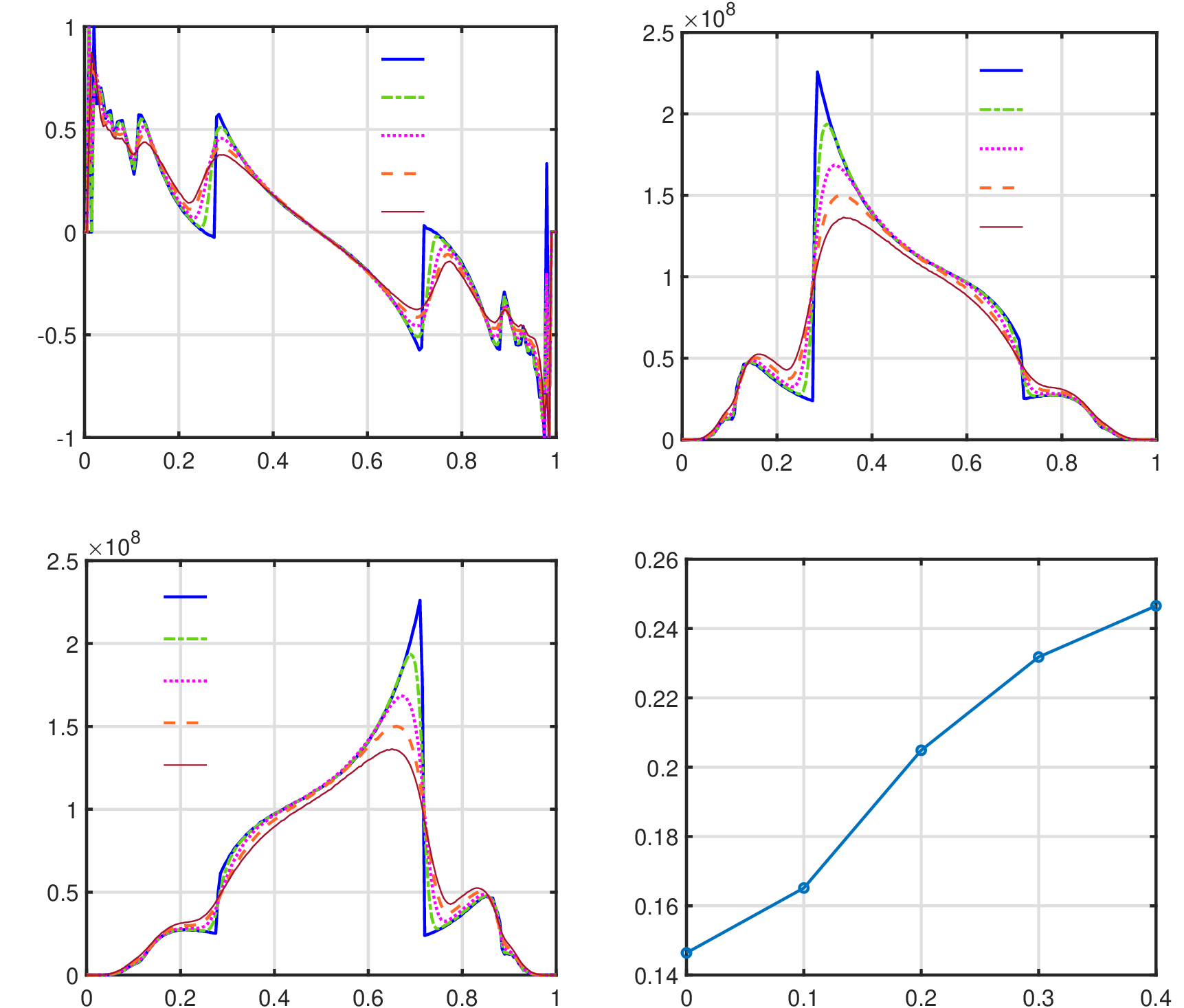}
     \begin{picture}(300,20)
         \put(26,203){{(a)}}
        \put(143,202){{(b)}}
         \put(26,97){{(c)}}
          \put(144,97){{(d)}}


        \put(91,207){\scriptsize ${\rho} = 0.0$}
        \put(91,199){\scriptsize ${\rho} = 0.1$}
        \put(91,191){\scriptsize ${\rho} = 0.2$}
        \put(91,184){\scriptsize ${\rho} = 0.3$}
        \put(91,177) {\scriptsize ${\rho} = 0.4$}
        
        \put(210,205){\scriptsize ${\rho} = 0.0$}
        \put(210,197){\scriptsize ${\rho} = 0.1$}
        \put(210,189){\scriptsize ${\rho} = 0.2$}
        \put(210,181){\scriptsize ${\rho} = 0.3$}
        \put(210,173) {\scriptsize ${\rho} = 0.4$}

        \put(48,100){\scriptsize ${\rho} = 0.0$}
        \put(48,92){\scriptsize ${\rho} = 0.1$}
        \put(48,84){\scriptsize ${\rho} = 0.2$}
        \put(48,75){\scriptsize ${\rho} = 0.3$}
        \put(48,67) {\scriptsize ${\rho} = 0.4$}


        \put(2,58){\rotatebox{90}{\small $dN/d\delta$}}  
         \put(123,163){\rotatebox{90}{\small $dN/d\delta$}}  
         \put(67,8){\small $\delta$}              
         \put(2,170){\rotatebox{90}{\small $S_z$}}         
          \put(186,116){\small $\delta$} 
         \put(122,63){\rotatebox{90}{\small $\langle S_z \rangle$}} 
        \put(186,8){\small ${\rho}$}                                  
    \end{picture}
\caption{(a) Energy‑resolved longitudinal spin polarization $S_z$ of the produced positrons as a function of the normalized energy fraction $\delta=\varepsilon_+/\omega$. The curves correspond to reduced squeezing parameters ${\rho}=0.0$ (blue), $0.1$ (green), $0.2$ (magenta), $0.3$ (orange), and $0.4$ (red).
(b,c) Density distributions of spin‑up (b) and spin‑down (c) positrons for different values of ${\rho}$. The spin‑up and spin‑down states are defined with respect to the $z$ axis.
(d) Average spin polarization $\langle S_z \rangle$, evaluated over the interval $0.15<\delta<0.27$, as a function of ${\rho}$.}
    \label{Fig:polarization}
\end{figure}

\subsection*{B. Positron polarization}

The influence of the squeezing parameter on the longitudinal polarization of the produced positrons is shown in Fig.~\ref{Fig:polarization}. The different curves correspond to different squeezing parameters. In general, low-energy positrons (\( \delta \lesssim 0.5 \)) inherit the laser helicity, whereas high-energy positrons (\( \delta \gtrsim 0.5 \)) exhibit helicity reversal. As \({\rho}\) increases, the polarization curve becomes smoother and undergoes discernible upward or downward shifts in several energy ranges [Fig.~\ref{Fig:polarization} (a)], indicating that squeezed light provides a handle to tune the spin polarization of positrons in addition to reshaping the spectrum.
Figures~\ref{Fig:polarization}(b) and (c) show the spin-resolved positron spectra \(dN/d\delta\). Spin-up positrons ($S_z>0$) dominate in the low-energy range (\(\delta \lesssim 0.5\)), where a pronounced harmonic structure appears; increasing \({\rho}\) suppresses these oscillations [Fig.~\ref{Fig:polarization} (b)]. 
Spin-down positrons  ($S_z<0$) dominate in the high-energy range (\(\delta \gtrsim 0.5\)), where the yield varies significantly with \({\rho}\), while the low-energy range remains nearly unchanged [Fig.~\ref{Fig:polarization} (c)]. 
Thus, changes in \(S_z\) at low energies are mainly due to spin-up positrons, whereas changes at high energies are driven by spin-down positrons.
To quantify the impact of the squeezing parameter on polarization, Fig.~\ref{Fig:polarization} (d) plots the average \(z\)-component of the positron polarization within \( \delta \in [0.15,\,0.27] \) versus \({\rho}\). The mean polarization rises monotonically from 0.14 for coherent light (${\rho}=0$) to 0.25 at \( {\rho} = 0.4 \),  demonstrating that squeezed-state driving can significantly modify spin-resolved pair production observables relative to the coherent-light case.

\subsection*{C. Positron angular distribution}

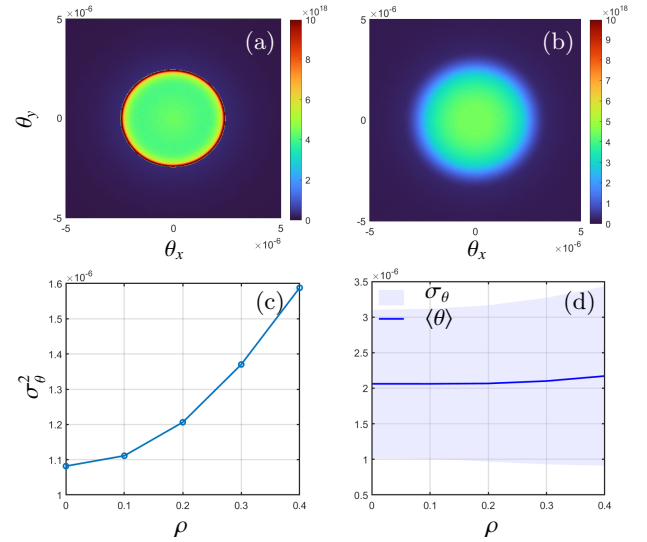
\begin{figure}
    \centering
    \includegraphics[width=0.49\textwidth]{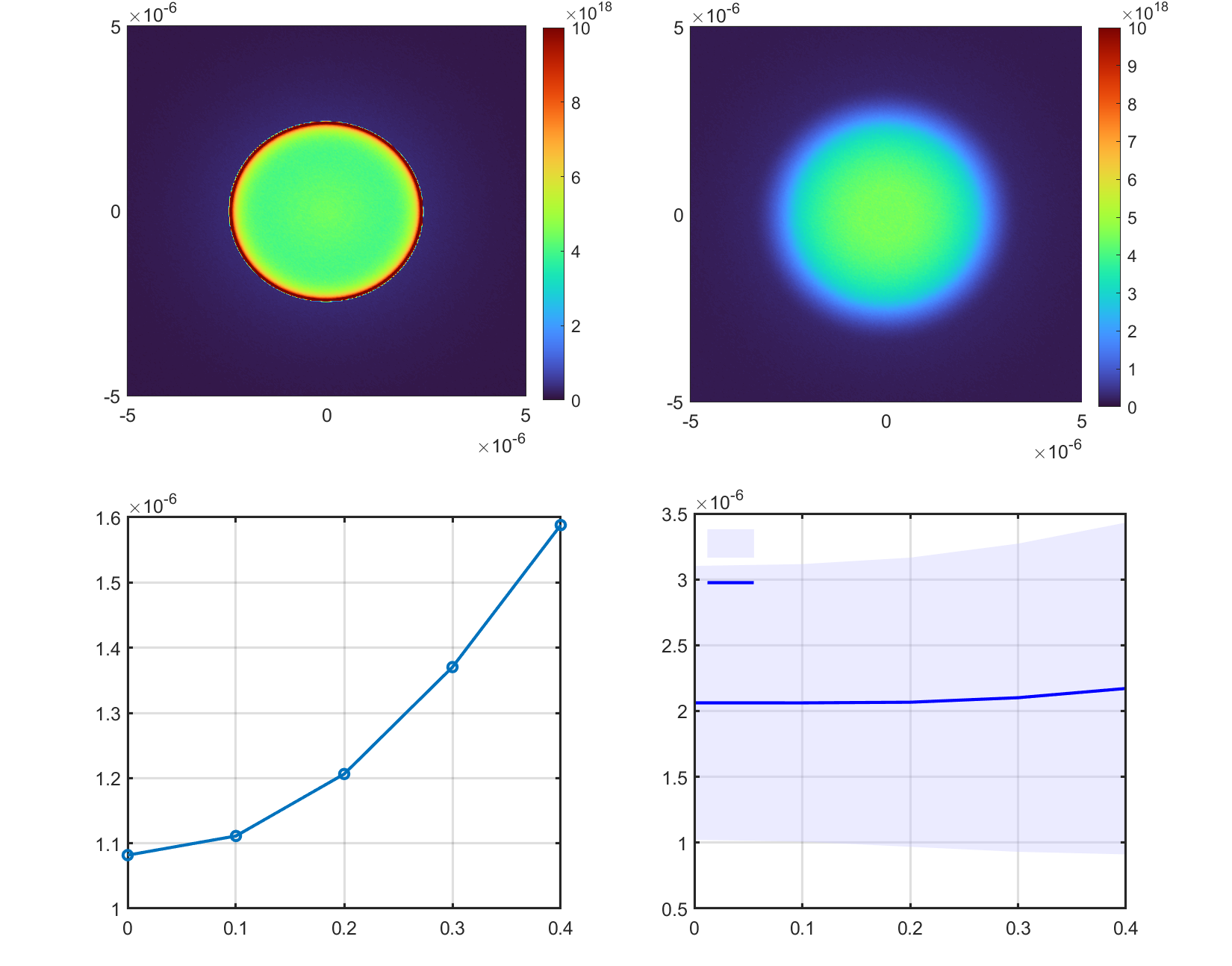}  
      \begin{picture}(0,0)

        \put(-30,190){\textcolor{white}{{(a)}}}
        \put(-116,160){\rotatebox{90}{\small $\theta_y$ }}
        \put(-60,112){\small $\theta_x$ }

        \put(82,190){\textcolor{white}{{(b)}}}
        \put(54,112){\small $\theta_x$ }

        \put(-26,92){{(c)}}
        \put(-115,58){\rotatebox{90}{\small $\sigma_{\theta}^2$ }}
        \put(-56,8){\small ${\rho}$}

        \put(90,92){{(d)}}
        \put(61,8){\small ${\rho}$}
        \put(38,86){\small $\langle \theta \rangle$}
        \put(38,96){\small $\sigma_{\theta}$}
            \end{picture}
\caption{Angular distributions of positrons, $d^2N/d\theta_x\,d\theta_y$ (rad$^{-2}$), as functions of the emission angles $\theta_x=\arctan(p_x/p_z)$ and $\theta_y=\arctan(p_y/p_z)$, for different reduced squeezing parameters: ${\rho}=0$ (a) and ${\rho}=0.4$ (b). 
(c) Polar‑angle variance $\sigma_{\theta}^2$ of the produced positrons as a function of the squeezing parameter ${\rho}$, characterizing the growth of angular spread under squeezed light. 
(d) Mean polar angle $\langle \theta \rangle$ (solid line) and the corresponding one‑sigma band $\langle \theta \rangle \pm \sigma_{\theta}$ (shaded region) versus ${\rho}$.}
    \label{fig:angular_dist}
\end{figure}

For the coherent-light case, the positron angular distribution is circularly symmetric and displays a hard-edged, ring-enhanced profile [Fig. \ref{fig:angular_dist} (a)]. This arises from the head-on geometry and the fixed field strength \(a_0\), which impose a sharp cutoff on the angular distribution:
$\theta\left(\delta,a_0\right)\lesssim \theta_{\max}\approx 2\times 10^{-6}~\text{rad}$ [Eq. (\ref{theta})]. 
Given the approximately uniform positron yield across different $\delta$ [Fig. \ref{fig:spectrum_nbw}], the angular distribution ($\theta\propto a_0/\delta$) remains nearly homogeneous within this kinematically allowed disk.

Introducing squeezing makes the field strength
\(a_0\) fluctuate from event to event.
Consequently, the kinematic limits are no longer fixed numbers
but are sampled from a distribution of cut-off values.
Averaging over this distribution blurs the sharp boundary of the
coherent case and fills in the central region, producing the broadened,
quasi-Gaussian pattern visible in
Fig.~\ref{fig:angular_dist}\,(b).
To quantify the broadening, we define the polar-angle variance
\begin{equation}
  \sigma_{\theta}^{2}
  =
  \langle\theta^{2}\rangle
  -
  \langle\theta\rangle^{2},
  \label{eq:sigtheta}
\end{equation}
where \(\langle\cdots\rangle\) denotes an average over the distribution $\tilde Q$.
Figure~\ref{fig:angular_dist}\,(c) plots \(\sigma_{\theta}^{2}\) versus
\({\rho}\).
The steady rise shows that the angular distribution widens as
squeezing grows.
This trend is driven mainly by fluctuations of the transverse momentum
\(p_{\perp}\): for small angles,
\(\theta\simeq p_{\perp}/p_{z}\), and the change of \(p_{z}\) with
\(a_{0}\) is suppressed by the large lepton energy
\(\varepsilon_{+}\), whereas \(p_{\perp}\propto a_{0}\) varies linearly
with the field strength.
Panel (d) complements the variance plot by showing the mean polar angle
\(\langle\theta\rangle\) (solid blue) together with its one-sigma
band \(\langle\theta\rangle\pm\sigma_{\theta}\)
(light-blue shading).
While the mean stays nearly constant, the widening band visually reflects the growth of \(\sigma_{\theta}\) seen in panel (c).
Taken together, panels (b)–(d) demonstrate that squeezing-induced fluctuations in the sampled field amplitude markedly enhance the angular divergence of the produced electron–positron pairs.

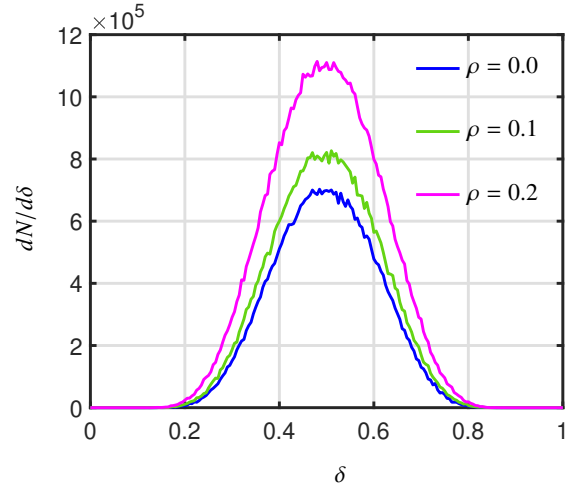
\begin{figure}
    \centering

    \includegraphics[width=0.45\textwidth]{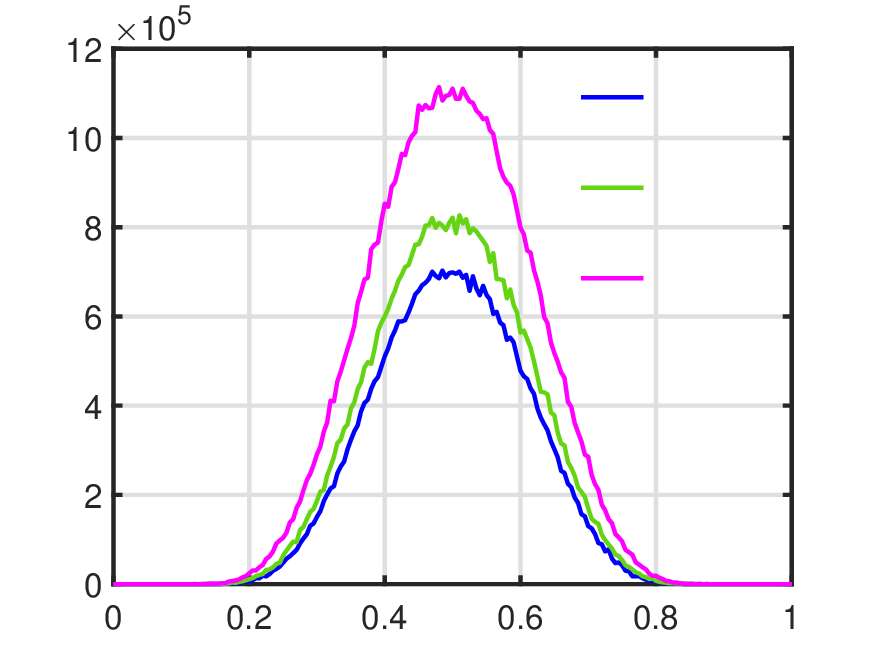}
   \begin{picture}(300,20)
       \put(180,166){${\rho} = 0.0$}
       \put(180,142){${\rho} = 0.1$}
       \put(180,118){${\rho} = 0.2$}
        \put(10,99){\rotatebox{90}{\small $dN/d\delta$}}  
        \put(130,12){\small $\delta$}              
    \end{picture}
    \caption{Energy spectrum of positrons generated in the NBW process for an initial photon energy of 16~GeV and a circularly polarized laser field with $a_0^2 = 2$, under different reduced squeezing parameters: ${\color{red}\rho} = 0.0$ (blue), $0.1$ (green), and $0.2$ (red).} 
    \label{fig:squeezing_high_a0}
\end{figure}

\subsection*{D. Impact of laser and $\gamma$‑photon parameters}

We next examine the NBW process in a head‑on collision between a $16~\mathrm{GeV}$ $\gamma$-ray beam ($N_\gamma=2\times 10^6$) and a circularly polarized laser pulse with intensity parameter $a_0^2=2$. Figure~\ref{fig:squeezing_high_a0} shows the resulting positron energy spectra for ${\rho}=0.0$, $0.1$, and $0.2$. In all three cases, the spectra are symmetric and broadly distributed around $\delta\simeq0.5$. Notably, the peak yield $dN/d\delta$ increases from approximately $6\times10^5$ at ${\rho}=0$ to $1.2\times10^6$ at ${\rho}=0.2$.

Compared with the setups in Secs.~III A--C, the present parameters imply a higher pair production threshold, requiring the absorption of approximately $n_{\min}\sim 32$ laser photons. For a fixed harmonic order $n$, energy-momentum conservation in head-on geometry
\(
k+n\kappa=q_+ + q_-,
\)
together with the quasi-momentum mass-shell conditions
\(
q_\pm^2=m_*^2,
\)
yields the allowed energy-fraction window~\cite{ritus1985quantum}
\[
\delta \in \bigl[\delta_-^{(n)},\;\delta_+^{(n)}\bigr],\qquad
\delta_\pm^{(n)}
   = \tfrac12\!\left[1 \pm
        \sqrt{1-\tfrac{n_{0}}{n}}\,
      \right],\numberthis
\]
whose width $\Delta\delta_n$ grows with~$n$ while the spacing between
adjacent channels shrinks. With increasing laser intensity, higher harmonic orders become accessible. For sufficiently large $n$,  the windows overlap so
strongly that the discrete harmonic pattern is washed out, and even a
coherent pulse already produces a smooth, quasi-continuous
spectrum; see Fig.~\ref{fig:squeezing_high_a0}.

When squeezed light is applied, shot-to-shot fluctuations in the field amplitude $a_0$ modify the NBW yield. Under the present parameters, a large number of high-order multiphoton channels contribute simultaneously. Although the probability of each individual channel depends only moderately on $a_0$, the broad statistical spread of $a_0$ introduced by the squeezed-state $Q$-function reweights the dense channel ensemble, leading to a nearly uniform, broadband enhancement of the differential spectrum $dN/d\delta$. The effect becomes more pronounced as  ${\rho}$ increases, since the broadened field-amplitude distribution preferentially amplifies the higher-order rates, which are more sensitive to large $a_0$ fluctuations.
By contrast, for previous parameters ($a_0^2 = 0.2$, $\omega_\gamma = 250~\mathrm{GeV}$, $\chi_\gamma \simeq 1.4$), only a limited number of low-order channels ($n \leq 5$) are kinematically accessible. In this case, the NBW spectrum retains distinct harmonic peaks, and the integrated yield approximately follows the perturbative scaling $W_n \propto a_0^{2n}$. Since squeezing-induced fluctuations affect only a small number of contributing terms, the overall yield is far less sensitive to changes in ${\rho}$ than in the case with many overlapping channels.
In summary, squeezing-induced intensity fluctuations have the greatest impact when the process involves a wide range of photon orders, where statistical reweighting can amplify high-order contributions. Conversely, their influence is relatively modest when the dynamics are dominated by a few well-separated channels.

\section{Conclusion}

We have investigated the nonlinear Breit--Wheeler process in intense squeezed light using a polarization-resolved Monte Carlo framework.
Within our approach, the pair production probability is obtained by averaging the strong-field QED rate over the squeezed-state field-amplitude distribution derived from the Husimi $Q$-function.
The observed modifications are most naturally interpreted as arising from the reweighting of field amplitudes induced by the quantum state of the source at fixed mean electric field amplitude.
As the squeezing parameter increases, the statistical weight of stronger-field realizations grows.
This leads to a systematic reshaping of the pair production signal: higher-order multiphoton channels are enhanced, harmonic structures are smoothed, and both energy and angular distributions broaden.
In the selected spectral window, the positron polarization degree also increases with the squeezing parameter, demonstrating that spin-resolved observables respond directly to the underlying field statistics.
A particularly notable effect is the suppression of the fundamental ($n=1$) photon-absorption channel.
Stronger-field realizations increase the dressed mass and thereby raise the kinematic threshold, so that the lowest-order channel becomes partially inaccessible, while higher-order channels remain open and acquire increased statistical weight.
These effects do not originate from a change in the average field strength or from a classical modification of the beam geometry.
Rather, they reflect how different quantum-state preparations of the driving field alter the distribution of field amplitudes entering the strong-field QED averaging procedure.
Our results therefore establish a direct theoretical connection between the quantum-optical statistics of the driving source and nonlinear Breit--Wheeler observables, and provide a framework for future studies exploring how the quantum state of intense light fields can influence strong-field particle production.

{\it Acknowledgements:}
This work is supported by the National Natural Science Foundation of China (Grants Nos.12474312 and 12574377), and the National Key R\&D Program of China (Grant No. 2021YFA1601700). 
 P.-L.~He acknowledges support from the Pujiang Program of the Shanghai Baiyulan Talent Plan (Grant No.~24PJA046), the Xiaomi Young Scholar Program, the Shanghai Jiao Tong University 2030 Initiative, and the Yangyang Development Fund.



    



\bibliography{prp}

\end{document}